\documentclass{svjour3}

\usepackage{graphics}
\usepackage{amssymb}
\usepackage{bm}% bold math
\usepackage{natbib}
\usepackage{fancyhdr}

\usepackage{latexsym}

\usepackage[abs]{overpic}

\usepackage{amsfonts}

\usepackage{amsmath}

\usepackage{mathrsfs}

\usepackage{graphicx}

\newcommand{\pp}[1]{\phantom{#1}}
\newcommand{\be}{\begin{eqnarray}}
\newcommand{\ee}{\end{eqnarray}}

\begin{document}

\title{
Imitating quantum probabilities: Beyond Bell's theorem and Tsirelson bounds
}
\author{Marek Czachor and Kamil Nalikowski}

\institute{Wydzia{\l} Fizyki Technicznej i Matematyki Stosowanej,
Politechnika Gda\'nska, 80-233 Gda\'nsk, Poland
}
\maketitle

\begin{abstract}
Local hidden-variable model of singlet-state correlations discussed in \citep{Czachor2021} is shown to be a particular case of an infinite hierarchy of local hidden-variable models based on an infinite hierarchy of calculi. Violation of Bell-type inequalities can be interpreted as a `confusion of languages' problem,  a result of mixing different but neighboring levels of the hierarchy. Mixing of non-neighboring levels  results in violations beyond the Tsirelson bounds.
\end{abstract}

\section{Introduction}

The question of whether specific values of physical quantities are revealed (and not created) by measurements is at the heart of the quantum measurement problem. A positive answer would mean that quantum mechanics is an incomplete theory, so  sub-quantum levels of reality are in principle possible
\citep{EPR,Bohm0,Bohm,Bell}. These deeper levels of reality could, at least theoretically, be used to hack quantum communication systems
\citep{BB,Ekert,BBM92,Gisin,hack0,hack}. 

The quantum hacking problem is similar but less ambitious than the quantum measurement one. 
Namely, it is enough if for a specific quantum system  we can find, at least in principle, a system with the same probabilities as for the quantum one, but in which the measurement results can be known in advance.  
If the imitation is meant as a possible means of attack on a quantum communication channel, we are not required to use one universal model for all quantum systems. As any hackers, we are free to modify the method of attack to modifications of the target. 
\footnote{The model of spin-1/2 proposed by \cite{Aerts1986} imitates a spin-1/2 particle in that it leads to correct probabilities. However, it cannot be regarded as an alternative representation of a spin-1/2 particle, as it does not involve quantum interference between probabilities, or change of sign of a spinor under $2\pi$ rotations. Still, it can be used in tests of hidden variable attacks on quantum communication channels. One concludes that the one-particle BB84 protocol is no longer secure, whereas the security of entangled-state two-particle protocols can be restored if one appropriately includes an additional random element \citep{ACP}. In principle, the non-Newtonian models we discuss in the present paper can be augmented by constructions of an Aerts type. Such a modification may be important for protocols that involve a sequence of measurements performed on a single qubit, where a hidden-variable representation of non-commutative measurements would be essential \citep{Czachor1992}. We will later see that measurements performed by Alice and Bob in the singlet state scenario remain commutative also in the non-Newtonian framework, and this explains why the model is local. This should be contrasted with the `rigid-rod' singlet-state generalization of the Aerts model, where the measurements performed by the two parties remain as non-commutative as in the single-particle case. The two-particle system corresponding to the `rigid rod'  Aerts model has to be regarded as an indivisible single entity, so is non-local in the sense of \cite{Bell}.}

The phrase `in principle' is here the keyword. 
In operational terms the question is if  one  can  {\it prove\/} the security of the communication system. 
Every proof is based on mathematical assumptions. The problem is if our assumptions are general enough.

A recent claim that there is an arithmetic (or calculus) fundamental  loophole in proofs of Bell's theorem creates a context for our paper \citep{Czachor2020}. A model that exactly reconstructs quantum probabilities typical of a two-particle singlet state has been recently constructed \citep{Czachor2021}. Quantum probabilities are there written in a Clauser-Horne local form \citep{CH}, detectors are perfect, observers have free will, hidden-variable probability density is non-negative. The difference is in the form of the integral used in the hidden-variable construction, and in the exact meaning of multiplication of functions under the integral. 

The integral is there {\it non-Newtonian\/} \citep{GK,BC}. In the literature it often appears under the name of {\it non-additive integral\/} \citep{Pap1993,Mesiar1995,Pap2002,G}. It is known to mathematicians since at least 1972, but was unknown to Bell in 1964, at the time of writing his classic paper. Its special cases are encountered in fuzzy calculus 
\citep{fuzzy calc,fuzzy dif}
and fractal analysis 
\citep{MC2015,ACK2016a,ACK2016b,ACK2018}. In particular, the Hausdorff integral on Koch fractals \citep{ES1,ES2,Czachor2019} is exactly non-Newtonian in this sense.

The multiplication is an operation from {\it projective arithmetic\/} \citep{Burgin77,Burgin2010,BC}. 
It can be regarded as an intermediate stage between ordinary multiplicaton and the tensor product.

The results are given by {\it projective bits\/}, 0 and 1, which are as ordinary real numbers as eigenvalues of projectors. Similarly to quantum bits, the projective bits corresponding to different projective arithmetics may be as incompatible as eigenvalues of non-commuting projectors. Yet, as opposed to quantum bits, the projective bits are  equipped with Einstein-Podolsky-Rosen type elements of reality.

`Arithmetic', often occurring in the plural form `arithmetics', is another important keyword here.

An arithmetic of real numbers consists of the set $\mathbb{R}$, supplemented by rules of addition and multiplication. 
Arithmetic is a language of mathematics. 
As there are different languages, there exist different arithmetics. Arithmetics can be either  isomorphic to one another or essentially different \citep{BC}.
The arithmetics we will discuss below are all isomorphic.

The fundamental theoretical freedom implicit in multitude of available arithmetics is not as valued by theoretical physicists as it deserves. 

We will argue  that quantum probabilities violate Bell-type inequalities in a similar sense as the decimal  inequality $0.1+0.1<1.0$ is `violated' by the binary equality $0.1+0.1=1.0$. Obviously, the problem disappears if we are cautious enough 
\footnote{The example involves a double `confusion of languages', occurring for both the arithmetic operation $+$ and the form of the numeral $0.1$. Numerals should be distinguished from numbers \citep{numeral}} and write,
$(0.1)_{10}+_{10} (0.1)_{10}<(1.0)_{10}$ and $(0.1)_{2}+_{2} (0.1)_{2}=(1.0)_{2}$.
Alternatively, the problem disappears if one is allowed to work exclusively with a single arithmetic. 

The problem also disappears if the arithmetic employed in theoretical calculations is exactly the same as the one that governs the rules of the physical world. However, there are arguments that problems with dark energy, dark matter, or Bell's theorem, may indicate that arithmetic is as physical as geometry
\citep{Entropy,MCdark2,B2002,B2005}.  In particular, systems at different scales may involve different arithmetic rules.

The goal of the present paper is to have a closer look at the Bell theorem example, quantum probabilities, and probabilities that are even far beyond the quantum ones. We will show that once we accept the possibility of a generalized calculus, not only are we capable of circumventing limitations of Bell's theorem, but we can violate the Tsirelson bounds characteristic of Hilbert space models of probability \citep{Tsirelson}. 
\footnote{
Similarly to \citep{Czachor2021} the construction has a status of a formal counterexample to Bell's theorem. We are not making any claims about universality of the proposed arithmetic.  All our results should be seen in the context of the hacking problem. In principle, any quantum communication protocol may require an attack based on a different arithmetic and calculus. Although we do not give any concrete physical example of a classical system that involves this concrete arithmetic, we are not aware of any argument for its non-existence. Formal similarity between non-Newtonian and fuzzy calculi suggests that examples should be looked for among fuzzy systems, as discussed many years ago by \cite{Pyk}. It seems there is no a priori reason to exclude non-Newtonian models of probability from discussions of security in cryptography, neither classical nor quantum.
}

\begin{figure}
\includegraphics[width=8 cm]{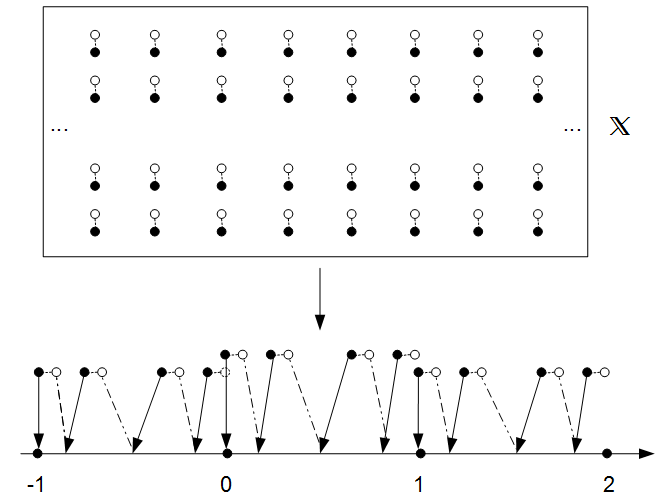}
\caption{$\mathbb{X}\subset\mathbb{R}^2$ consists of countably many right-open Cantor sets placed next to one another. In order to construct the bijection $f_\mathbb{X}$ one first moves a $k$th Cantor set into $[k,k+1)\subset\mathbb{R}$. Now, any point of the resulting subset of $\mathbb{R}$ can be parametrized by a ternary number whose digits involve only 0 and 2. In the final step one replaces 2 by 1, and treats the resulting sequence of bits as a binary representation of a point in $\mathbb{R}$. This step is geometrically equivalent to gluing together the appropriate black and white points, so that the holes of the Cantor set disappear. The construction is reversible: One first cuts and splits according to a fixed recipe, creating Cantor sets. In order to perform arithmetic operations in $\mathbb{X}$, one first maps its elements into $\mathbb{R}$, then performs the ordinary operations in $\mathbb{R}$, and finally moves the result back into $\mathbb{X}$. }
\label{Fig X}
\end{figure}

\section{Arithmetic elements of reality}

In order to have a feel of a possible real-world context of our considerations, think of a real number as an activity pattern of some brain or artificial neural network. The set of such patterns, let us call it $\mathbb{X}$, must be in a one-to-one relation with $\mathbb{R}$ (i.e. has cardinality of the continuum), but otherwise can be quite weird (of a fractal type, say). We are interested in formulating arithmetic, calculus and, ultimately, physics and cryptography in a setting involving `strange representations' of reals. The resulting formalism is known in mathematics under the names of projective arithmetics and non-Newtonian calculi.

Now, consider two sets,  $\mathbb{X}$ and $\mathbb{Y}$, with cardinalities of the continuum $\mathbb{R}$. Just to focus our attention, let $\mathbb{X}\subset\mathbb{R}^2$ be the countable collection of Cantor sets depicted in Fig.~\ref{Fig X}, while $\mathbb{Y}=\mathbb{R}_+$. What is essential, because of identical cardinalities of $\mathbb{X}$, $\mathbb{Y}$,  and $\mathbb{R}$, there exist bijections $f_\mathbb{X}:\mathbb{X}\to \mathbb{R}$ and
$f_\mathbb{Y}:\mathbb{Y}\to \mathbb{R}$ (for example, $f_\mathbb{Y}(y)=\ln y$; a construction of $f_\mathbb{X}$ is shown in Fig.~\ref{Fig X}). The bijections induce in $\mathbb{X}$ and $\mathbb{Y}$ 
the four aritmetic operations,
\be
x\oplus_\mathbb{X} y  &=& f_\mathbb{X}^{-1}\big( f_\mathbb{X}(x)+f_\mathbb{X}(y)\big),\label{X1}\\
x\ominus_\mathbb{X} y  &=& f_\mathbb{X}^{-1}\big( f_\mathbb{X}(x)-f_\mathbb{X}(y)\big),\label{X2}\\
x\odot_\mathbb{X} y  &=& f_\mathbb{X}^{-1}\big( f_\mathbb{X}(x)\cdot f_\mathbb{X}(y)\big),\label{X3}\\
x\oslash_\mathbb{X} y  &=& f_\mathbb{X}^{-1}\big( f_\mathbb{X}(x)/f_\mathbb{X}(y)\big),\label{X4}
\ee
and 
\be
x\oplus_\mathbb{Y} y  &=& f_\mathbb{Y}^{-1}\big( f_\mathbb{Y}(x)+f_\mathbb{Y}(y)\big),\label{Y1}\\
x\ominus_\mathbb{Y} y  &=& f_\mathbb{Y}^{-1}\big( f_\mathbb{Y}(x)-f_\mathbb{Y}(y)\big),\label{Y2}\\
x\odot_\mathbb{Y} y  &=& f_\mathbb{Y}^{-1}\big( f_\mathbb{Y}(x)\cdot f_\mathbb{Y}(y)\big),\label{Y3}\\
x\oslash_\mathbb{Y} y  &=& f_\mathbb{Y}^{-1}\big( f_\mathbb{Y}(x)/f_\mathbb{Y}(y)\big).\label{Y4}
\ee
The arithmetics given by (\ref{X1})--(\ref{Y4}) are called projective. The neutral elements, $0_\mathbb{X}=f_\mathbb{X}^{-1}(0)$ (projective zero in $\mathbb{X}$),
$1_\mathbb{X}=f_\mathbb{X}^{-1}(1)$ (projective one in $\mathbb{X}$), $0_\mathbb{Y}=f_\mathbb{X}^{-1}(0)$ (projective zero in $\mathbb{Y}$),
$1_\mathbb{Y}=f_\mathbb{Y}^{-1}(1)$ (projective one in $\mathbb{Y}$), can be termed the projective bits. 

Of course, $x\oplus_\mathbb{X} 0_\mathbb{X}=x$, $x\odot_\mathbb{X} 1_\mathbb{X}=x$, $y\oplus_\mathbb{Y} 0_\mathbb{Y}=y$, 
$y\odot_\mathbb{Y} 1_\mathbb{Y}=y$, for any $x\in \mathbb{X}$,  $y\in \mathbb{Y}$. The projective bits corresponding to different arithmetics are as incompatible as qubits corresponding to incompatible observables in quantum mechanics. In particular, expresions such as $1_\mathbb{X}+1_\mathbb{Y}$, 
$1_\mathbb{X}\oplus_\mathbb{X} 1_\mathbb{Y}$,
$1_\mathbb{X}\oplus_\mathbb{Y}1_\mathbb{Y}$,
or the like, are meaningless (try to add $1_\mathbb{X}=f_\mathbb{X}^{-1}(1)=(1,0)\in\mathbb{X}\subset\mathbb{R}^2$ to 
$1_\mathbb{Y}=f_\mathbb{Y}^{-1}(1)=e\in\mathbb{R}_+$), whereas 
$1_\mathbb{X}\oplus_\mathbb{X} 1_\mathbb{X}=2_\mathbb{X}$ or
$1_\mathbb{Y}\oplus_\mathbb{Y}1_\mathbb{Y}=2_\mathbb{Y}$,
make perfect sense.

Yet, there is  an intriguing difference between projective bits and quantum bits. Namely, in spite of their incompatibility, the projective bits posses common `elements of reality'. For example, both $1_\mathbb{X}=f_\mathbb{X}^{-1}(1)$ and $1_\mathbb{Y}=f_\mathbb{Y}^{-1}(1)$ are images of the same element $1\in\mathbb{R}$. There exists an analogue of an entangled state that correlates $0_\mathbb{X}$ with $0_\mathbb{Y}$, and 
$1_\mathbb{X}$ with $1_\mathbb{Y}$. It is easy to invent such a state, just think of a correlation via their common elements or reality: $0,1\in\mathbb{R}$.

 This is the reason why probabilistic models involving projective bits may look quantum, and yet be equipped with Einstein-Podolsky-Rosen-type elements of reality. In such models one can prove various forms of Bell inequalities, but one cannot prove Bell's theorem. 
Models based on projective arithmetics can easily mimic quantum probabilities. 
Cryptographic protocols based on projective bits may look as secure as their quantum counterparts. However, any attempt of proving their fundamental security will be hopeless.

\section{Real numbers and their elements of reality}

In this section we will construct a concrete but infinite hierarchy of reals $\mathbb{X}=R^k$, $k=0,\pm1,\pm2,\dots$ Our goal is to construct an infinite hierarchy of projective bits and models of probability associated with $R^k$. Different levels of the hierarchy will play the roles of `macroscopic' (i.e. observer, or human) , `microscopic' (i.e. quantum), `sub-microscopic' (i.e. sub-quantum, or hidden-variable) levels of `reality'. 

Let us denote by $R^k$ the set of reals equipped with four arithmetic operations, $\oplus^k$, $\ominus^k$, $\odot^k$, $\oslash^k$, constructed as follows. Let $f: \mathbb{R}\to \mathbb{R}$ be an increasing bijection with inverse $g=f^{-1}$. 
(Example: $f(x)=x^3$, $g(x)=\sqrt[3]{x}$.)

Denote by $g^k$ the $k$th-fold composition of $g$ with itself, 
$g^k=g\circ\dots\circ g$ ($k$ times). Analogously, $g^{-k}=f\circ \dots \circ f=g^{-1}\circ \dots \circ g^{-1}$. We accept the convention that  $g^1=g$ and $g^0(x)=x$.
Repeating the general paradigm sketched in the previous section, we can write the following formulas,
\be
x\oplus^k y  &=& g^k\Big( g^{-k}(x)+g^{-k}(y)\Big)=g^k\Big( f^k(x)+f^k(y)\Big),\\
x\ominus^k y  &=& g^k\Big( g^{-k}(x)-g^{-k}(y)\Big)=g^k\Big( f^k(x)-f^k(y)\Big),\\
x\odot^k y  &=& g^k\Big( g^{-k}(x)\cdot g^{-k}(y)\Big)=g^k\Big( f^k(x)\cdot f^k(y)\Big),\\
x\oslash^k y  &=& g^k\Big( g^{-k}(x)/g^{-k}(y)\Big)=g^k\Big( f^k(x)/f^k(y)\Big).
\ee
(For $f(x)=x^3$, $g(x)=\sqrt[3]{x}$, the multiplication is unchanged, but $x\oplus^0 y=x+y$, $x\oplus^1 y=\sqrt[3]{x^3+y^3}$, $x\oplus^{-1}
y=(\sqrt[3]{x}+\sqrt[3]{y})^3$, $x\oplus^2 y=\sqrt[9]{x^9+y^9}$, etc.)

An arithmetic is a set equipped with four arithmetic operations and ordering relation $\le$. The latter symbol will be skipped, so
$R^k=\{\mathbb{R},\oplus^k, \ominus^k, \odot^k, \oslash^k\}$,
\be
R^0=\{\mathbb{R},\oplus^0, \ominus^0, \odot^0, \oslash^0\}=
\{\mathbb{R},+, -, \cdot, /\}.
\ee
The set $\mathbb{R}$ (or the arithmetic $R^0$) plays the role of arithmetic elements of reality for $R^k$. We assume that $f$ is increasing so that  the ordering relation $\le$ is independent of $k$. 

It is obvious that:
\be
g^{-k}(x\oplus^k y)  &=& g^{-k}(x)+g^{-k}(y),\\
g^{-k}(x\ominus^k y)  &=& g^{-k}(x)-g^{-k}(y),\\
g^{-k}(x\odot^k y)  &=& g^{-k}(x)\cdot g^{-k}(y),\\
g^{-k}(x\oslash^k y)  &=&  g^{-k}(x)/g^{-k}(y).
\ee
Accordingly, the bijection $f^k=g^{-k}$ is simultaneously an isomorphism of arithmetics,
\be
f^k:R^k \to R^0.
\ee
This immediately implies that multiplication and addition in
$R^k$ are commutative and associative, and multiplication is distributive with respect to addition. Neutral elements of addition and multiplication in 
$R^k$ are, respectively, $0_k=g^k(0)$ and $1_k=g^k(1)$. In what follows we will often assume $0_k=0$ and $1_k=1$, but this is not essential. It is sometimes even useful to have a more general case at our disposal.

An {\it arithmetic Copernican principle\/} follows from the following observation: The four arithmetic operations, for any $k,l\in\mathbb{Z}$, are related by
\be
x\odot^{k+l} y
&=&
g^l\Big(
g^{-l}(x)\odot^k g^{-l}(y)\Big)
=
g^k\Big(
g^{-k}(x)\odot^l g^{-k}(y)\Big),\\
x\oslash^{k+l} y
&=&
g^l\Big(
g^{-l}(x)\oslash^k g^{-l}(y)\Big)
=
g^k\Big(
g^{-k}(x)\oslash^l g^{-k}(y)\Big),\\
x\oplus^{k+l} y
&=&
g^l\Big(
g^{-l}(x)\oplus^k g^{-l}(y)\Big)
=
g^k\Big(
g^{-k}(x)\oplus^l g^{-k}(y)\Big),\\
x\ominus^{k+l} y
&=&
g^l\Big(
g^{-l}(x)\ominus^k g^{-l}(y)\Big)
=
g^k\Big(
g^{-k}(x)\ominus^l g^{-k}(y)\Big).
\ee
Equivalently,
\be
g^{-k}
\left(
x\odot^{k+l} y
\right)
&=&
g^{-k}(x)\odot^l g^{-k}(y),\\
g^{-k}
\left(
x\oslash^{k+l} y
\right)
&=&
g^{-k}(x)\oslash^l g^{-k}(y),\\
g^{-k}
\left(
x\oplus^{k+l} y
\right)
&=&
g^{-k}(x)\oplus^l g^{-k}(y),\\
g^{-k}
\left(
x\ominus^{k+l} y
\right)
&=&
g^{-k}(x)\ominus^l g^{-k}(y).
\ee
The bijection
$f^k=g^{-k}$ is simultaneously an isomorphism of $R^{k+l}$ and $R^l$,
\be
f^k:R^{k+l}\to R^l.
\ee
The {\it Copernican principle\/} states that the role of a 0th level can be taken over by any $l$. In fact, there is nothing wrong with the notation where
\be
R^l=\{\mathbb{R},\oplus^l, \ominus^l, \odot^l, \oslash^l\}=
\{\mathbb{R},+, -, \cdot, /\}.
\ee
There is absolutely nothing special about $l=0$. Accordingly, any $R^l$ can be regarded as `the' ordinary arithmetic we are taught at elementary schools.
(For $f(x)=x^3$, $g(x)=\sqrt[3]{x}$, we could write  $x+ y=\sqrt[3]{x^3\oplus^{-1}y^3}$.)

Physics should be formulated in a way that does not depend on $l$. Equivalently, physics should be formulated for all $l$ simultaneously. 

One can go further and formulate an arithmetic principle of relativity: No physical experiment can detect a privileged $l$. However,  a relative $l$ might be observable.

Copernican and relativistic arithmetic principles are inherited from arithmetic by calculus.

\section{A hierarchy of non-Newtonian calculi}

A hierarchy of arithmetics leads to a hierarchy of calculi. 
In the terminology of Grossman and Katz \citep{GK} the calculi are non-Newtonian. In particular, a derivative of a function $A:\mathbb{R}\to \mathbb{R}$ reads
\be
\frac{{\rm D}_l A(x)}{{\rm D}_k x}
&=&
\lim_{\delta\to 0}
\Big(
A(x\oplus^k\delta_k)\ominus^l A(x)
\Big)
\oslash^l \delta_l,\label{D/Dx}
\ee
where $R^k$ is the arithmetic of the domain, and $R^l$ is the arithmetic of the codomain.
Here $\delta_k=g^k(\delta)$, $\delta_l=g^l(\delta)$, and we assume continuity of $g^k$ and $g^l$ at $x=0\in\mathbb{R}$ (otherwise $0_k$ and $0_l$ would be ambiguous). The derivative is linear and satisfies 
an appropriate Leibniz rule,
\be
\frac{{\rm D}_l \big(A(x)\oplus^l B(x)\big)}{{\rm D}_k x}
&=&
\frac{{\rm D}_l A(x)}{{\rm D}_k x}
\oplus^l
\frac{{\rm D}_l B(x)}{{\rm D}_k x},\\
\frac{{\rm D}_l \big(A(x)\odot^l B(x)\big)}{{\rm D}_k x}
&=&
\left(
\frac{{\rm D}_l A(x)}{{\rm D}_k x}\odot^l B(x)
\right)
\oplus^l
\left(
A(x)\odot^l
\frac{{\rm D}_l B(x)}{{\rm D}_k x}
\right).
\ee
As we can see, there is no difficulty with differentiating functions whose domains and codomains belong to different levels of the hierarchy.

Integration of $A:\mathbb{R}\to \mathbb{R}$ is defined in a way that guarantees the two fundamental theorems of calculus (under standard assumptions about differentiability and  continuity):
\be
\int_a^b
\frac{{\rm D}_l A(x)}{{\rm D}_k x} {\rm D}_k x
&=&
A(b)\ominus^l A(a),\\
\frac{{\rm D}_l }{{\rm D}_k x}
\int_a^x
A(y) {\rm D}_k y
&=&
A(x).
\ee
In order to make the above formulas less abstract, consider the following commutative diagram, written in three equivalent forms, each of them stressing a different aspect of the calculus:
\be
\begin{array}{rcl}
\mathbb{R}                & \stackrel{A}{\longrightarrow}       & \mathbb{R}               \\
f^k{\Big\downarrow}   &                                     & {\Big\downarrow}f^l  \\
\mathbb{R}                & \stackrel{\tilde A}{\longrightarrow}   & \mathbb{R}
\end{array},
\quad
\begin{array}{rcl}
\mathbb{R}                & \stackrel{A}{\longrightarrow}       & \mathbb{R}               \\
f^k{\Big\downarrow}    &                                     & {\Big\uparrow}g^l  \\
\mathbb{R}                & \stackrel{\tilde A}{\longrightarrow}   & \mathbb{R}
\end{array},
\quad
\begin{array}{rcl}
\mathbb{R}                & \stackrel{A}{\longrightarrow}       & \mathbb{R}               \\
g^k{\Big\uparrow}    &                                     & {\Big\downarrow}f^l  \\
\mathbb{R}                & \stackrel{\tilde A}{\longrightarrow}   & \mathbb{R}
\end{array}.
\label{diagramf}
\ee
The left form shows that $A$ can be regarded as a function between arithmetics, $A:R^k\to R^l$. 
The middle form shows that $A$ is defined by means of an intermediate function $\tilde A: R^0\to R^0$, subject to the ordinary rules of the calculus we have learned at schools. 
The right form is an explicit definition of $\tilde A$ in terms of the bijections and the function $A$,
\be
\tilde A(r)
&=&
f^l\Big(A\big(g^k(r)\big)\Big).
\ee
The latter leads to a very simple and useful form of the derivative (\ref{D/Dx}),
\be
\frac{{\rm D}_l A(x)}{{\rm D}_k x}
&=&
g^l
\left(
\frac{{\rm d}\tilde A\big(f^k(x)\big)}{{\rm d} f^k(x)}
\right)
\ee
where ${\rm d}\tilde A(r)/{\rm d}r$ is the usual Newtonian derivative we know from the standard course of calculus in $R^0$. Notice that the bijection itself, $g^l$, is not differentiated, apparently violating the chain rule for compositions of functions. A deeper analysis shows, however, that it is in fact differentiated, but  by means of a non-Newtonian chain rule \citep{BC,Czachor2019}. Moreover, the bijections $f_\mathbb{X}:\mathbb{X}\to R^0$ and 
$f_\mathbb{Y}:\mathbb{Y}\to R^0$ are  always non-Newtonian differentiable, even if they are Newtonian non-differentiable (think 
of the Cantor set example from Fig.~\ref{Fig X}).

The associated integral reads,
\be
\int_a^b
A(x) {\rm D}_k x
&=&
g^l
\left(
\int_{f^k(a)}^{f^k(b)}
\tilde A(r) {\rm d}r
\right).\label{integr}
\ee
Here ${\rm d}r$ denotes the usual (Riemann, Lebesgue, etc.) integral in $R^0$.

\section{Some useful properties of the integral}

In order to get used to non-Newtonian derivatives and integrals, let us explicitly verify four important properties. 

\medskip

\noindent
(a) The fundamental theorem of calculus follows from the following calculation:
\be
\frac{{\rm D}_l}{{\rm D}_k x}
\int_a^x
A(y) {\rm D}_k y
&=&
\frac{{\rm D}_l}{{\rm D}_k x}
g^l
\left(
\int_{f^k(a)}^{f^k(x)}
\tilde A(r) {\rm d}r
\right)
=
g^l
\left(
\frac{{\rm d}}{{\rm d} f^k(x)}
\int_{f^k(a)}^{f^k(x)}
\tilde A(r) {\rm d}r
\right)\nonumber\\
&=&
g^l
\left(
\tilde A\big(f^k(x)\big)
\right)
=A(x),
\ee
\be
\int_a^b
\frac{{\rm D}_l A(x)}{{\rm D}_k x} {\rm D}_k x
&=&
g^l
\left(
\int_{f^k(a)}^{f^k(b)}
\frac{{\rm d}\tilde A(r)}{{\rm d}r} {\rm d}r
\right)
=
g^l
\left(
\tilde A\big(f^k(b)\big)-\tilde A\big(f^k(a)\big)
\right)
\\
&=&
g^l
\left(
f^l\circ g^l\circ\tilde A\circ f^k(b)-f^l\circ g^l\circ\tilde A\circ f^k(a)
\right)
=
A(b)\ominus^l A(a).
\ee 
(b) Additivity of the integral is an important ingredient of proofs of Bell-type inequalities. Its non-Newtonian form reads,
\be
\int_a^b
\big(A(x)\oplus^l B(x)\big) {\rm D}_k x
&=&
\int_a^b
A(x) {\rm D}_k x
\oplus^l
\int_a^b
 B(x) {\rm D}_k x.\label{property b}
\ee
Note that additivity is given with respect to $R^l$, but not with respect to $R^0$. 
Integrals possessing such a generalized form of additivity are called non-additive and are ubiquitously present in fuzzy calculus. 

In order to prove the (non-)additivity, we have to find the explicit form of $\widetilde{A\oplus^l B}(r)$, defined by the diagram
\be
\begin{array}{rcl}
\mathbb{R}                & \stackrel{A\oplus^l B}{\longrightarrow}       & \mathbb{R}               \\
g^k{\Big\uparrow}   &                                     & {\Big\downarrow}f^l  \\
\mathbb{R}                & \stackrel{\widetilde{A\oplus^l B}}{\longrightarrow}   & \mathbb{R}
\end{array},
\ee
because
\be
\int_a^b
\big(A(x)\oplus^l B(x)\big) {\rm D}_k x
=
\int_a^b
(A\oplus^l B)(x) {\rm D}_k x
=
g^l
\left(
\int_{f^k(a)}^{f^k(b)}
\widetilde{A\oplus^l B}(r) {\rm d}r
\right).\label{integr+}
\ee
Let us compute,
\be
\widetilde{A\oplus^l B}(r)
&=&
f^l\Big((A\oplus^l B)\big(g^k(r)\big)\Big)
=
f^l\Big(A\big(g^k(r)\big)\oplus^l B\big(g^k(r)\big)\Big)
\\
&=&
f^l\circ g^l\Big(f^l\circ A\circ g^k(r)+ f^l\circ B\circ g^k(r)\Big)=
\tilde A(r)+\tilde B(r).
\ee
Accordingly,
\be
\int_a^b
\big(A(x)\oplus^l B(x)\big) {\rm D}_k x
&=&
g^l
\left(
\int_{f^k(a)}^{f^k(b)}
\tilde A(r){\rm d}r
+
\int_{f^k(a)}^{f^k(b)}
\tilde B(r){\rm d}r
\right)
\\
&=&
g^l
\left[
f^l\circ g^l
\left(
\int_{f^k(a)}^{f^k(b)}
\tilde A(r){\rm d}r
\right)
+
f^l\circ g^l
\left(
\int_{f^k(a)}^{f^k(b)}
\tilde B(r){\rm d}r
\right)
\right]
\nonumber
\\
&=&
\int_a^b
A(x) {\rm D}_k x
\oplus^l
\int_a^b
 B(x) {\rm D}_k x
.\label{integr+1}
\ee
(c) In the context of Clauser-Horne local-realistic formulas we will often encounter integrals of $\odot^l$ products of functions. We will now  show that
\be
\int_a^b
\big(A(x)\odot^l B(x)\big) {\rm D}_k x
&=&
g^l
\left(
\int_{f^k(a)}^{f^k(b)}
\tilde A(r)\tilde B(r) {\rm d}r
\right).\label{integr .}\label{property c}
\ee
The proof is similar to the one for sums. 
In order to prove it, we have to find the explicit form of $\widetilde{A\odot^l B}(r)$, defined by the diagram
\be
\begin{array}{rcl}
\mathbb{R}                & \stackrel{A\odot^l B}{\longrightarrow}       & \mathbb{R}               \\
g^k{\Big\uparrow}   &                                     & {\Big\downarrow}f^l  \\
\mathbb{R}                & \stackrel{\widetilde{A\odot^l B}}{\longrightarrow}   & \mathbb{R}
\end{array},
\ee
because
\be
\int_a^b
\big(A(x)\odot^l B(x)\big) {\rm D}_k x
=
\int_a^b
(A\odot^l B)(x) {\rm D}_k x
=
g^l
\left(
\int_{f^k(a)}^{f^k(b)}
\widetilde{A\odot^l B}(r) {\rm d}r
\right).\label{integr.1}
\ee
Let us compute,
\be
\widetilde{A\odot^l B}(r)
&=&
f^l\Big((A\odot^l B)\big(g^k(r)\big)\Big)
=
f^l\Big(A\big(g^k(r)\big)\odot^l B\big(g^k(r)\big)\Big)
\\
&=&
f^l\circ g^l\Big(f^l\circ A\circ g^k(r)\cdot f^l\circ B\circ g^k(r)\Big)=
\tilde A(r)\tilde B(r).
\ee
As a by-product of the calculation we observe that for $\tilde B(r)=\tilde C=$~constant, we find
\be
\int_a^b
\big(A(x)\odot^l C\big) {\rm D}_k x
=
C\odot^l\int_a^b
A(x) {\rm D}_k x,
\ee
where
\be
C=
g^l\Big(\tilde B\big(f^k(x)\big)\Big)=g^l(\tilde C).
\ee
(d) Bayes rules for conditional probabilities (both Newtonian and non-Newtonian) involve ratios of probabilities. We will often need the formula
\be
g^l
\left(
\frac{\int_{f^k(a)}^{f^k(b)}\tilde A(r) {\rm d}r}{\int_{f^k(a)}^{f^k(b)}\tilde B(r) {\rm d}r}
\right)
&=&
g^l
\left(
\frac{f^l\circ g^l\left(\int_{f^k(a)}^{f^k(b)}\tilde A(r) {\rm d}r\right)}{f^l\circ g^l\left(\int_{f^k(a)}^{f^k(b)}\tilde B(r) {\rm d}r\right)}
\right)
\nonumber\\
&=&
\left(
\int_a^b
A(x)  {\rm D}_k x
\right)
\oslash^l
\left(
\int_a^b
B(x)  {\rm D}_k x
\right).\label{property d}
\ee

\section{Singlet state probabilities...}

In what follows, we will concentrate on two-spin-1/2 singlet-state probabilities. The same probabilities can be obtained by means of two Mach-Zehnder interferometers. In the spin-1/2 example, the parameter denoted below by $\alpha$ or $\beta$ represents an angle of rotation of a Stern-Gerlach device. In the Mach-Zehnder case the parameters represent interferometric phases.

We are interested in: probabilities of conditions,
\be
P_{0_1} &=& P_{1_1} =\langle\psi|\hat P_{0_1}\otimes I|\psi\rangle=\langle\psi|\hat P_{1_1}\otimes I|\psi\rangle=\frac{1}{2},\\
P_{0_2} &=& P_{1_2} =\langle\psi|I\otimes \hat P_{0_2}|\psi\rangle=\langle\psi|I\otimes \hat P_{1_2}|\psi\rangle=\frac{1}{2},
\ee
and joint probabilities,
\be
P_{0_10_2} &=& P_{1_11_2} 
=\langle\psi|\hat P_{0_1}\otimes \hat P_{0_2}|\psi\rangle
=\langle\psi|\hat P_{1_1}\otimes \hat P_{1_2}|\psi\rangle=\frac{1}{2}\sin^2\frac{\alpha-\beta}{2},\\
P_{0_11_2} &=& P_{1_10_2} 
=\langle\psi|\hat P_{0_1}\otimes \hat P_{1_2}|\psi\rangle
=\langle\psi|\hat P_{1_1}\otimes \hat P_{0_2}|\psi\rangle=\frac{1}{2}\cos^2\frac{\alpha-\beta}{2}.
\ee

\section{...and how to imitate them}
\label{Section fake}

The problem is to find a bijection $g$ (and its corresponding arithmetic and calculus) such that 
\be
P_{A_1A_2} &=& \int \chi_{A_1}(x)\odot^l \chi_{A_2}(x)\odot^l \rho(x){\rm D}_lx,\label{eq 49},\\
P_{A} &=& \int \chi_{A}(x)\odot^l \rho(x){\rm D}_lx = \frac{1}{2},
\ee
where the $\chi$s are projective-bits valued characteristic functions, with $0_l=0$ and $1_l=1$. The latter is guaranteed by $g(0)=0$, $g(1)=1$. 
The $\chi$s so defined behave as orthogonal projectors,
\be
1 &=&\chi_0(x)\oplus^l \chi_1(x),\\
0 &=& \chi_0(x)\odot^l \chi_1(x),\\
\chi_0(x) &=& \chi_0(x)\odot^l \chi_0(x),\\
\chi_1(x) &=& \chi_1(x)\odot^l \chi_1(x).
\ee
In other words, we want to reconstruct quantum probabilities in spite of the fact that elements of reality do exist.

The right-hand side of (\ref{eq 49}) implies
\be
P_{0_10_2}\oplus^l
P_{0_11_2}\oplus^l
P_{1_10_2}\oplus^l
P_{1_11_2}
=
1.\label{eq 51}
\ee
However, if for hacking purposes we want to fake quantum probabilities, the following equality must be also true (for the same probabilities!)
\be
P_{0_10_2}+
P_{0_11_2}+
P_{1_10_2}+
P_{1_11_2}
=
1.\label{eq 53}
\ee
Indeed, an observer will test the theory by means of the simplest, frequentist definition of probability, so will divide and add in the ordinary way, consistent 
with (\ref{eq 53}), and not with (\ref{eq 51}).

Finding a nontrivial bijection $g$ that simultaneously guarantees (\ref{eq 51}) and (\ref{eq 53}),  is not an entirely trivial task, but its solution exists.

Notice that the coexistence of  (\ref{eq 51}) and (\ref{eq 53}),  is just another aspect of the Copernican principle. 
How to recognize which of the two additions, $+$ or $\oplus^l$,  is `the ordinary' one?

If we change the notation, 
\be
\oplus^l\leftrightarrow  +, \quad\odot^l\leftrightarrow \cdot,
\ee 
we obtain the celebrated Clauser-Horne local-realistic form of singlet-state probabilities, 
\be
P_{A_1A_2} &=& \int \chi_{A_1}(x)\cdot  \chi_{A_2}(x)\cdot  \rho(x){\rm D}_lx,\label{eq 49'}
\ee
normalized with respect to both (\ref{eq 51}) and (\ref{eq 53}), 
a formula which is widely believed to be impossible.

A construction of $g$  boils down to the following two lemmas.

\medskip

\noindent
{\bf Lemma 1.} 
$g(p)+g(1-p)=1$ for any $p\in[0,1]$  if and only if
\be
g(p)=\frac{1}{2} + h\left(p-\frac{1}{2}\right),\label{g2}
\ee
where  $h(-x)=-h(x)$. Any such $g$ has a fixed point at $p=1/2$.

\medskip

\noindent
{\bf Lemma 2.} Consider four joint probabilities $p_{0_10_2}$, $p_{1_11_2}$, $p_{0_11_2}$, $p_{1_10_2}$, satisfying 
\be
\sum_{AB}p_{AB} &=& 1,\label{L2a}\\
\sum_{A}p_{AA_2} &=& \sum_{A}p_{A_1 A}=\frac{1}{2}.\label{L2b}
\ee
A sufficient condition for
\be
\sum_{AB}g(p_{AB}) &=& 1,\label{L2G}
\ee
is given by $g(p)=\frac{1}{2}G(2p)$, where $G$ satisfies Lemma~1. Any such $g$ has a fixed point at $p=1/4$. $\Box$

\medskip

\noindent
{\it Proofs\/}. Lemma~1 is discussed in detail in \citep{Entropy,Czachor2021}.  
In the context of spin-1/2 measurements,  (\ref{L2b}) is equivalent to rotational invariance. 
(\ref{L2b}) implies $\sum_{A_1}2p_{A_1A_2}=1$. So, by Lemma~1,
\be
G(2p_{0_10_2})+G(2p_{1_10_2}) =1=G(2p_{0_11_2})+G(2p_{1_11_2}),
\ee
and thus
\be
G(2p_{0_10_2})+G(2p_{1_10_2})+G(2p_{0_11_2})+G(2p_{1_11_2})
=2,
\ee
which proves (\ref{L2G}) for $g(p)=\frac{1}{2}G(2p)$. Finally, $g(1/4)=\frac{1}{2}G(1/2)=1/4$.
An example of $g$ is shown in Fig.~\ref{Figfg}.

\section{Two Copernican lemmas}

The first lemma states that if two numbers  are binary probabilities at one level of the Copernican hierarchy, the same two numbers  play the same role at any level of the hierarchy.

\medskip

\noindent
{\bf Lemma 3.} ({\it Binary probabilities\/}) Assume $g:[0,1]\to [0,1]$ is invertible and 
\be
g(p)+g(1-p)=1, \quad \textrm{for any $p\in[0,1]$}.
\ee
Then 
\be
g^{k}(p)\oplus^l g^{k}(1-p)=1_l, \quad \textrm{for any $p\in[0,1]$ and any $k,l\in\mathbb{Z}$}.
\ee
In particular, if $g(1)=1$ then 
\be
g^{k}(p)\oplus^l g^{k}(1-p)=1, \quad \textrm{for any $p\in[0,1]$ and any $k,l\in\mathbb{Z}$}.
\ee
\medskip

\noindent
{\it Proof\/}. 
We first prove that 
\be
g^{k}(p)+ g^{k}(1-p)=1, \quad \textrm{for any $p\in[0,1]$ and any $k\in\mathbb{Z}$}.
\ee
It is enough to prove that 
\be
g^{k}(p)
&=&
\frac{1}{2}+h\left(p-\frac{1}{2}\right)
\ee 
for some antisymmetric $h$. 
Now, consider two functions described by Lemma~1, $g_1,g_2:[0,1]\to [0,1]$,
\be
g_1(p) &=& \frac{1}{2} + h_1\left(p-\frac{1}{2}\right), \label{g_1}\\
g_2(p) &=& \frac{1}{2} + h_2\left(p-\frac{1}{2}\right). \label{g_2}
\ee
Then,
\be
g_1\circ g_2(p) 
&=&
\frac{1}{2} + h_1\left(g_2(p)-\frac{1}{2}\right)\\
&=&
\frac{1}{2} + h_1\left(
\frac{1}{2} + h_2\left(p-\frac{1}{2}\right)
-\frac{1}{2}\right)\\
&=&
\frac{1}{2} + h_1\left(
 h_2\left(p-\frac{1}{2}\right)
\right)\\
&=&
\frac{1}{2} + h_1\circ
 h_2\left(p-\frac{1}{2}
\right).
\ee
As $h_1$ and $h_2$ are antisymmetric, their composition is antisymmetric as well,
\be
h_1\circ h_2(-x)
&=&
h_1\big(h_2(-x)\big)
=
h_1\big(-h_2(x)\big)=
-h_1\big(h_2(x)\big)
=
-h_1\circ h_2(x).
\ee
Accordingly, $g_{12}=g_1\circ g_2$ also satisfies assumptions of Lemma~1 with $h_{12}=h_1\circ h_2$.
Beginning with $g_1=g_2=g$, we prove by induction that
\be
g^{k}(p)
&=&
\frac{1}{2}+h^k\left(p-\frac{1}{2}\right),
\ee 
for any  non-negative $k\in \mathbb{Z}$.

For negative $k$ one first notices that
an invertible $g:[0,1]\to [0,1]$ satisfies 
\be
g(1-p)+g(p)
&=&
1
\ee
if and only if 
\be
g^{-1}(1-p)+g^{-1}(p)
&=&
1.
\ee
Indeed, let $g(1-p)+g(p)=1$. Then 
\be
p' &=& g(p)=\frac{1}{2} + h\left(p-\frac{1}{2}\right), \\
p' -\frac{1}{2}&=&  h\left(p-\frac{1}{2}\right), \\
h^{-1}\left(p' -\frac{1}{2}\right)&=&  p-\frac{1}{2}, \\
\frac{1}{2}+h^{-1}\left(p' -\frac{1}{2}\right)&=&  p=g^{-1}(p').
\ee
$h(-x)=-h(x)$ implies $h^{-1}(-x)=-h^{-1}(x)$. Hence, $g^{-1}$ satisfies assumptions of Lemma~1, and thus
$g^{-1}(1-p)+g^{-1}(p)=1$. Accordingly, $g(1-p)+g(p)=1$ implies $g^{-1}(1-p)+g^{-1}(p)=1$. 

Now, let $g^{-1}(1-p)+g^{-1}(p)=1$. Then, by what we have just proved,
\be
1=(g^{-1})^{-1}(1-p)+(g^{-1})^{-1}(p)=g(1-p)+g(p).
\ee
Beginning with $g_1=g_2=g^{-1}$, we prove by induction that
\be
g^{-k}(p)
&=&
\frac{1}{2}+h^{-k}\left(p-\frac{1}{2}\right).
\ee 
Finally, replacing in the above formulas $k$ by $k-l$, we get
\be
g^{k-l}(p)+ g^{k-l}(1-p)=1=g^{-l}\big(g^{k}(p)\big)+ g^{-l}\big(g^{k}(1-p)\big), \quad \textrm{for any $k,l\in\mathbb{Z}$}.
\ee
Acting on both sides of the latter equality with $g^l$ we end the proof.$\Box$
\medskip

The second lemma states that if four numbers are 2-bit rotationally invariant probabilities at one level of the Copernican hierarchy, the same four numbers  play the same role at any level of the hierarchy.

\medskip

\noindent
{\bf Lemma 4.} ({\it 2-bit rotationally symmetric probabilities\/}) 
Consider four joint probabilities $p_{0_10_2}$, $p_{1_11_2}$, $p_{0_11_2}$, $p_{1_10_2}$, and a bijection $g:\mathbb{R}\to\mathbb{R}$ from Lemma~2. Then
\be
g^k(p_{0_10_2})\oplus^l g^k(p_{1_10_2})\oplus^l g^k(p_{0_11_2})\oplus^l g^k(p_{1_11_2})=1_l=g^l(1),
\quad \textrm{for any $k,l\in\mathbb{Z}$.}
\ee
In particular, if $g(1)=1$ then 
\be
g^k(p_{0_10_2})\oplus^l g^k(p_{1_10_2})\oplus^l g^k(p_{0_11_2})\oplus^l g^k(p_{1_11_2})=1, 
\quad \textrm{for any $k,l\in\mathbb{Z}$.}
\ee\medskip

\noindent
{\it Proof\/}. 
By the preceding lemma, if
\be
G(p)+G(1-p)=1, \quad \textrm{for any $p\in[0,1]$},
\ee
then 
\be
G^{k}(p)+G^{k}(1-p)=1, \quad \textrm{for any $p\in[0,1]$ and any $k\in\mathbb{Z}$}.
\ee
Next, we  note that $g^k(p)=\frac{1}{2}G^k(2p)$, for any $k\in\mathbb{Z}$. 
Indeed,
$g^0(p)=p=\frac{1}{2}(2p)=\frac{1}{2}G^0(2p)$.
$g^1(p)=\frac{1}{2}G^1(2p)$ by definitions of $g^1=g$ and $G^1=G$. For $k=2$,
\be
g^2(p)=g\big(g(p)\big)=\frac{1}{2}G\big(2g(p)\big)=\frac{1}{2}G\left(2\frac{1}{2}G(2p)\right)
=
\frac{1}{2}G^2(2p).
\ee
We continue by induction for any positive $k\in\mathbb{Z}$. For negative, $k\in\mathbb{Z}$, it remains to show that  $g^{-1}(p)=\frac{1}{2}G^{-1}(2p)$ and then continue by induction. 
As $p'=g(p)=\frac{1}{2}G(2p)$, then 
\be
\frac{1}{2}G^{-1}(2p')=p=g^{-1}(p').
\ee
This ends the proof that $g^k(p)=\frac{1}{2}G^k(2p)$, for any $k\in\mathbb{Z}$.   
Note that $g^k(1/4)=\frac{1}{2}G^k(1/2)=1/4$ because $1/2$ is a fixed point of $G$, and thus of $G^k$ for any $k$. 
But now, by Lemma~2, we find
\be
1=\sum_{AB}g^{k-l}(p_{AB})=\sum_{AB}g^{-l}\left(g^{k}(p_{AB})\right)
\ee
Acting on both sides of the latter equality with $g^l$ we end the proof. $\Box$

\medskip

\noindent
{\bf Remark.} 
The fact that 
\be
g^{k}(p)\oplus^l g^{k}(1-p)=g^{k}(p)+ g^{k}(1-p)
\ee
does not yet imply that any linear combination of probabilities is independent of $l$. For example, typically
\be
g^{k}(p)\ominus^l g^{k}(1-p)\neq g^{k}(p)- g^{k}(1-p),
\ee
unless $g^l(x)=x$ for any $x$. General linear combinations of probabilities, such as those occurring in Bell-type inequalities, 
in general do depend on the choice of $l$. 

\medskip

\section{Singlet-state probabilities in a Clauser-Horne form}

The above two lemmas are used to determine the form of calculus that will lead to a hierarchy of hidden-variable models which include quantum singlet-state probabilities as a special case.

\begin{figure}
\includegraphics[width=8 cm]{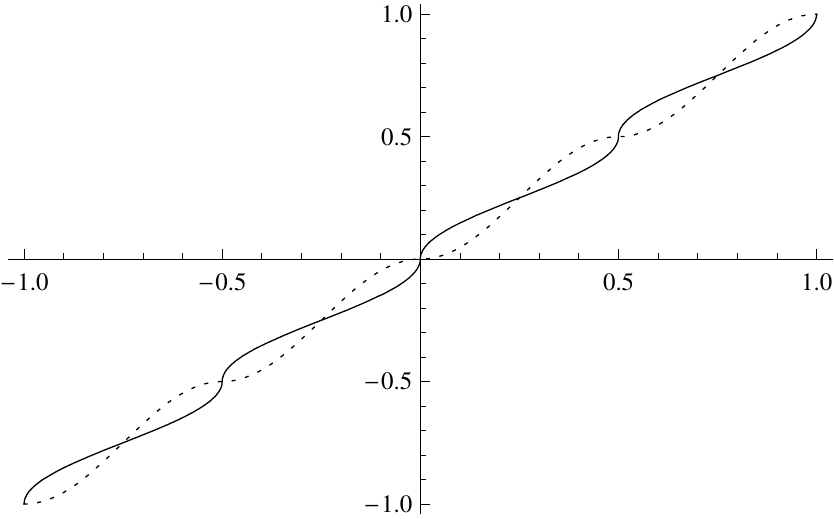}
\caption{One-to-one $f:\mathbb{R}\to \mathbb{R}$  (full) and its inverse $g$ (dotted) defined by  (\ref{f})  and (\ref{f^-1}). $g(x)$ is antisymmetric and satisfies $g(x)=g(x+1/2)-1/2=h(x)$ occurring in Lemma~1.}
\label{Figfg}
\end{figure}

To this end, consider the following bijection $f:\mathbb{R}\to\mathbb{R}$, and its inverse $g$,
\be
f(x) &=& \frac{n}{2}+\frac{1}{\pi}\arcsin\sqrt{2x -n}, \label{f}\\
g(x) &=&\frac{n}{2}+ \frac{1}{2}\sin^2\pi \left(x-\frac{n}{2}\right), \label{f^-1}\\
&\pp=&
\textrm{for $\frac{n}{2}\le x\le \frac{n+1}{2}$, $n\in\mathbb{Z}$}.
\ee
Their restriction to $[0,1]$ satisfies Lemmas 1 and 2. 
Actually, the antisymmetric function $h(x)$ from Lemma~1 is here equal to $g(x)$ itself,
\be
h(x)=g\left(x+\frac{1}{2}\right)-\frac{1}{2}=g(x).
\ee
$g$ and $f$ have infinitely many fixed points (see Fig.~\ref{Figfg}),
\be
g(n/4) = n/4 = f(n/4), \quad \textrm{for  $n\in\mathbb{Z}.$}
\ee
Let us introduce the following characteristic functions $\tilde\chi_A$ on a unit circle. Here $r\in[0,2\pi)$ is the polar angle, and $\alpha,\beta\in[0,\pi)$, $\alpha\ge\beta$:
\be
\tilde\chi_{0_1}(r) &=& \tilde\chi_{+\alpha}(r)
=
\left\{
\begin{array}{cl}
1 & \textrm{for $r\in[\alpha,\alpha+\pi)$},\\
0 & \textrm{for $r\not\in[\alpha,\alpha+\pi)$},
\end{array}
\right.
\\
\tilde\chi_{1_1}(r)
&=&
\tilde\chi_{-\alpha}(r)
=
1- \tilde\chi_{+\alpha}(r),\\
\tilde\chi_{0_2}(r) &=& \tilde\chi_{+\beta}(r)
=
\left\{
\begin{array}{cl}
1 & \textrm{for $r\in[\beta,\beta+\pi)$},\\
0 & \textrm{for $r\not\in[\beta,\beta+\pi)$},
\end{array}
\right.
\\
\tilde\chi_{1_2}(r)
&=&
\tilde\chi_{-\beta}(r)
=
1- \tilde\chi_{+\beta}(r).
\ee
The characteristic functions map $R^0$ into $R^0$. They define new characteristic functions $\chi_A$ at the 1st level of the Copernican hierarchy,
\be
\begin{array}{rcl}
\mathbb{R}                & \stackrel{\chi_A}{\longrightarrow}       & \mathbb{R}               \\
f{\Big\downarrow}    &                                     & {\Big\uparrow}g  \\
\mathbb{R}                & \stackrel{\tilde \chi_A}{\longrightarrow}   & \mathbb{R}
\end{array}.
\ee
Probability density at the 1st level of the hierarchy is defined by the diagram,
\be
\begin{array}{rcl}
\mathbb{R}                & \stackrel{\rho}{\longrightarrow}       & \mathbb{R}               \\
f{\Big\downarrow}    &                                     & {\Big\uparrow}g  \\
\mathbb{R}                & \stackrel{\tilde \rho}{\longrightarrow}   & \mathbb{R}
\end{array}.
\ee
It is normalized,
\be
1=\int_{g(0)}^{g(2\pi)}\rho(x){\rm D}_1 x=g\left(\int_0^{2\pi}\tilde\rho(r){\rm d}r\right)
\ee
With our choice of $g$ we find $g(0)=0$, $g(1)=1$, $g(2\pi)=(2\pi)'=6.30175$ (note that $(2\pi)'>2\pi=6.28319$), and $\tilde\rho(r)=1/(2\pi)$. 
Property (\ref{property c}) implies that 
\be
\int_{0}^{(2\pi)'}\chi_{0_1}(x)\odot^1\rho(x){\rm D}_1 x
&=&g\left(\int_0^{2\pi}\tilde\chi_{0_1}(r)\tilde\rho(r){\rm d}r\right)
=
g\left(\frac{1}{2\pi}\int_\alpha^{\alpha+\pi}{\rm d}r\right)
\nonumber\\
&=&
g\left(\frac{1}{2}\right)=\frac{1}{2},
\ee
and analogously with the remaining single-bit probabilities. Joint probabilities read
\be
\int_{0}^{(2\pi)'}\chi_{0_1}(x)\odot^1\chi_{0_2}(x)\odot^1\rho(x){\rm D}_1 x
&=&g\left(\int_0^{2\pi}\tilde\chi_{0_1}(r)\tilde\chi_{0_2}(r)\tilde\rho(r){\rm d}r\right)
\nonumber\\
&=&
g\left(\frac{1}{2\pi}\int_\beta^\alpha{\rm d}r\right)
=
g\left(\frac{\alpha-\beta}{2\pi}\right)
\nonumber\\
&=&
\frac{1}{2}\sin^2 \left(\pi\frac{\alpha-\beta}{2\pi}\right)
=
\frac{1}{2}\sin^2 \frac{\alpha-\beta}{2}.
\ee
One analogously proves,
\be
\int_{0}^{(2\pi)'}\chi_{0_1}(x)\odot^1\chi_{1_2}(x)\odot^1\rho(x){\rm D}_1 x
&=&g\left(\int_0^{2\pi}\tilde\chi_{0_1}(r)\tilde\chi_{1_2}(r)\tilde\rho(r){\rm d}r\right)
\\
&=&g\left(\int_0^{2\pi}\tilde\chi_{0_1}(r)\Big(1-\tilde\chi_{0_2}(r)\Big)\tilde\rho(r){\rm d}r\right)
\\
&=&
g\left(\frac{1}{2}-\frac{\alpha-\beta}{2\pi}\right)
=
\frac{1}{2}\sin^2 \left(\frac{\pi}{2}-\frac{\alpha-\beta}{2}\right)
\nonumber\\
&=&
\frac{1}{2}\cos^2 \frac{\alpha-\beta}{2},
\ee
and the remaining two singlet-state probabilities.

The four probabilities indeed sum to 1 in two different ways. The standard, `macroscopic' way reads,
\be
\frac{1}{2}\cos^2 \frac{\alpha-\beta}{2}
+
\frac{1}{2}\sin^2 \frac{\alpha-\beta}{2}
+
\frac{1}{2}\sin^2 \frac{\alpha-\beta}{2}
+
\frac{1}{2}\cos^2 \frac{\alpha-\beta}{2}
=1.\label{trig}
\ee
At the level of the $\oplus^1$ addition the calculation looks as follows,
\be
1&=&
g\left(\frac{1}{2}-\frac{\alpha-\beta}{2\pi}\right)
\oplus^1
g\left(\frac{\alpha-\beta}{2\pi}\right)
\oplus^1
g\left(\frac{\alpha-\beta}{2\pi}\right)
\oplus^1
g\left(\frac{1}{2}-\frac{\alpha-\beta}{2\pi}\right)\\
&=&
g\left(\frac{1}{2}-\frac{\alpha-\beta}{2\pi}+\frac{\alpha-\beta}{2\pi}+\frac{\alpha-\beta}{2\pi}+\frac{1}{2}-\frac{\alpha-\beta}{2\pi}\right)
=g(1).\label{fixed}
\ee
Technically speaking, (\ref{trig}) is a property of trigonometric functions, while (\ref{fixed}) is a property of fixed points of the bijection $g$. 
It is clear that the latter will hold for a much larger class of $g$'s, not only the ones corresponding to quantum probabilities.

\section{Violation of Tsirelson bounds for Clauser-Horne inequality}

The standard Clauser-Horne inequality for classical probabilities (in this section we employ the original notation from the Clauser and Horne paper), fulfilling the spherical symmetry condition $p_{1}^{j}(a')=p_{2}^{k}(b)=1/2$, reads
\be
0\le 
p_{12}^{jk}(a,b)
-
p_{12}^{jk}(a,b')
+
p_{12}^{jk}(a',b)
+
p_{12}^{jk}(a',b')
\le 1.
\ee
It is not satisfied by quantum probabilities. The latter are subject to the Tsirelson bounds \footnote{Tsirelson is the most popular spelling of this name, but the original paper spelled the name as Cirel'son.}
\be
-\frac{\sqrt{2}-1}{2}
 &=&-0.20711\nonumber\\
&\le&
p_{12}^{++}(a,b)
-
p_{12}^{++}(a,b')
+
p_{12}^{++}(a',b)
+
p_{12}^{++}(a',b')
\nonumber\\
&\le& 
\frac{\sqrt{2}+1}{2}=1.20711.
\ee
Spherically symmetric joint probabilities cannot be greater than $1/2$, therefore the absolute bounds for the Clauser-Horne combination of probabilities are
\be
-0.5
\le
p_{12}^{++}(a,b)
-
p_{12}^{++}(a,b')
+
p_{12}^{++}(a',b)
+
p_{12}^{++}(a',b')
\le
1.5
\ee
Now, assuming that the Clauser-Horne  inequality is true at the 0th level of the Copernican hierarchy defined by a bijection $g$, whose fixed points are located at least at 0, $\frac{1}{4}$, $\frac{1}{2}$,  and 1, 
\be
0\le 
p_{12}^{jk}(a,b)
\ominus^0
p_{12}^{jk}(a,b')
\oplus^0
p_{12}^{jk}(a',b)
\oplus^0
p_{12}^{jk}(a',b')
\le 1,
\ee
we can always prove  an $l$th-level analogue of the Clauser-Horne inequality,
\be
0
\le
g^l\left(p_{12}^{++}(a,b)\right)
\ominus^l
g^l\left(p_{12}^{++}(a,b')\right)
\oplus^l
g^l\left(p_{12}^{++}(a',b)\right)
\oplus^l
g^l\left(p_{12}^{++}(a',b')\right)
\le 
1.\nonumber\\
\label{CHl}
\ee
Such an inequality cannot be violated, because it can be proved.

The question is, what can we say about the possible left and right bounds for the same probabilities if we employ the `standard' arithmetic of the macroscopic observer, 
\be
?\,
\le
g^l\left(p_{12}^{++}(a,b)\right)
-
g^l\left(p_{12}^{++}(a,b')\right)
+
g^l\left(p_{12}^{++}(a',b)\right)
+
g^l\left(p_{12}^{++}(a',b')\right)
\le 
\, ?
\label{CH?}
\ee
The left and right Clauser-Horne  bounds from (\ref{CHl}) are no longer valid for (\ref{CH?}), unless by $\pm$ one means $\oplus^l$ and $\ominus^l$. 

\begin{figure}
\includegraphics[width=8 cm]{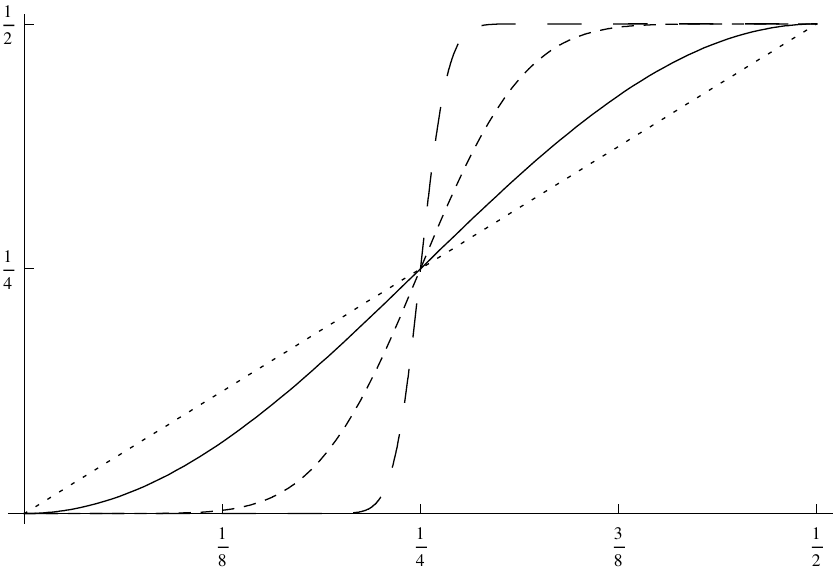}
\caption{The essence of the proof of (\ref{lim 05}) can be understood  if one looks at the plots of $g^0$ (dotted), $g^1$ (full),  $g^3$ (dashed), and $g^6$ (long-dashed). Our example  corresponds to $\theta=\pi/4$, but this concrete value  is irrelevant. In fact,  $\lim_{l\to\infty}g^l(x)=0$, for any $x\in(0,1/4)$, and 
$\lim_{l\to\infty}g^l(x)=0.5$, for any $x\in(1/4,1/2)$.}
\label{Fig gl}
\end{figure}

Starting with the same probabilities that were used to construct the non-Newtonian model of singlet state correlations, let us assume 
\be
p_{12}^{++}(a,b) &=& \frac{|\alpha-\beta|}{2\pi}=\frac{\theta}{2\pi},\\
p_{12}^{++}(a,b') &=& \frac{|\alpha-\beta'|}{2\pi}=\frac{3\theta}{2\pi},\\
p_{12}^{++}(a',b) &=& \frac{|\alpha'-\beta|}{2\pi}=\frac{\theta}{2\pi},\\
p_{12}^{++}(a',b') &=& \frac{|\alpha'-\beta'|}{2\pi}=\frac{\theta}{2\pi},
\ee
where $0\le 3\theta<\pi$, the expression we have to estimate is
\be
X(g^l,\theta)
=
3g^l\left(\frac{\theta}{2\pi}\right)
-
g^l\left(\frac{3\theta}{2\pi}\right)
\ee
The singlet-state example corresponds in this range of $x$ to $g^1(x)=\frac{1}{2}\sin^2 (\pi x)$. For  $\theta=\pi/4$ we then find
\be
X(g^1,\pi/4)
&=&
\frac{3}{2}\sin^2 \left(\pi \frac{\pi/4}{2\pi}\right)
-
\frac{1}{2}\sin^2 \left(\pi \frac{3\pi/4}{2\pi}\right)
\nonumber\\
&=&
\frac{3}{2}\sin^2 \frac{\pi}{8}
-
\frac{1}{2}\sin^2 \frac{3\pi}{8}=-0.20711.
\ee
For any $\theta$ satisfying $0<\frac{\theta}{2\pi}<\frac{1}{4}$ and 
$\frac{1}{4}<\frac{3\theta}{2\pi}<\frac{1}{2}$ one can prove that 
\be
\lim_{l\to\infty}X(g^l,\theta)=-0.5\label{lim 05}
\ee
if $g^1(x)=\frac{1}{2}\sin^2 (\pi x)$. The essence of the proof is the same as of Banach's fixed point theorem,
and consists of two limits,
\be
\lim_{l\to\infty}g^l\left(\frac{\theta}{2\pi}\right) &=0& \label{lim 0},\\
\lim_{l\to\infty}g^l\left(\frac{3\theta}{2\pi}\right) &=& 0.5\label{lim 1/2},
\ee
as illustrated in Fig.~\ref{Fig gl}.

\begin{figure}
\includegraphics[width=8 cm]{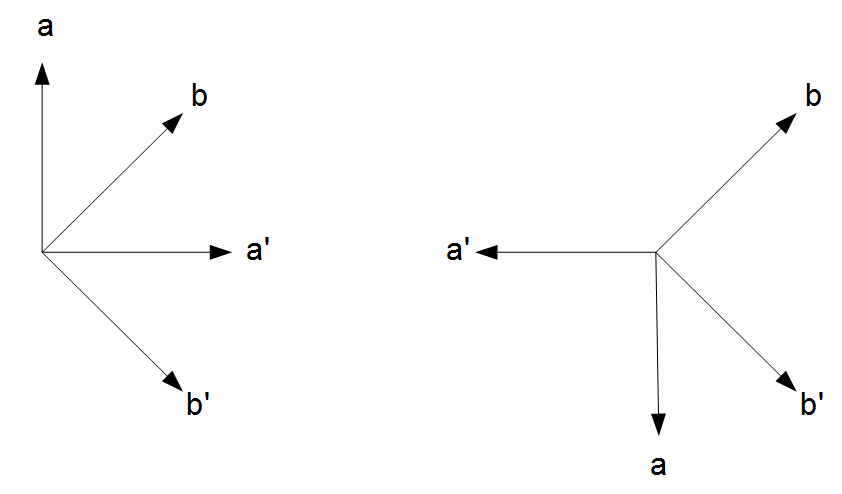}
\caption{Left: The configuration of spin directions that leads to maximal left violations (\ref{Xmin0})--(\ref{Xmin5}). Right: The configuration of spin directions that leads to maximal right violations (\ref{Xmax0})--(\ref{Xmax5}).}
\label{Fig4}
\end{figure}

The convergence is fast:
\be
X(g^0,\pi/4) &=& 0,\label{Xmin0}\\
X(g^1,\pi/4) &=& -0.20711,\\
X(g^2,\pi/4) &=& -0.39602,\\
X(g^3,\pi/4) &=& -0.48669,\\
X(g^4,\pi/4) &=& -0.49978,\\
&\vdots& \nonumber\\
X(g^\infty,\pi/4) &=& -0.5.\label{Xmin5}
\ee
Figure~\ref{Fig4} shows two configurations of Stern-Gerlach devices associated with the Clauser-Horne linear combination (\ref{CH?}). The left configuration guarantees the maximal left bounds (\ref{Xmin0})--(\ref{Xmin5}), the right one implies the right bounds (\ref{Xmax0})--(\ref{Xmax5}). Indeed, changing directions of $a$ and $a'$ we rather get the following four joint probabilities,
\be
p_{12}^{++}(a,b) &=& \frac{|\alpha-\beta|}{2\pi}=\frac{3\theta}{2\pi},\\
p_{12}^{++}(a,b') &=& \frac{|\alpha-\beta'|}{2\pi}=\frac{\theta}{2\pi},\\
p_{12}^{++}(a',b) &=& \frac{|\alpha'-\beta|}{2\pi}=\frac{3\theta}{2\pi},\\
p_{12}^{++}(a',b') &=& \frac{|\alpha'-\beta'|}{2\pi}=\frac{3\theta}{2\pi},
\ee
with $\theta=\pi/4$. Accordingly,
\be
X(g^l,\theta)
=
3g^l\left(\frac{3\theta}{2\pi}\right)
-
g^l\left(\frac{\theta}{2\pi}\right),
\ee
and
\be
X(g^0,\pi/4) &=& 1,\label{Xmax0}\\
X(g^1,\pi/4) &=& 1.20711,\\
X(g^2,\pi/4) &=& 1.39602,\\
X(g^3,\pi/4) &=& 1.48669,\\
X(g^4,\pi/4) &=& 1.49978,\\
&\vdots& \nonumber\\
X(g^\infty,\pi/4) &=& 1.5.\label{Xmax5}
\ee

\section{Conclusions}

Violation of Bell-type inequalities occurs if one combines $k$-th level probabilities by means of arithmetic rules that are valid for some other level $l$. When interpreted in this way, violation of Bell's inequality is no longer very mysterious. A sentence formulated in one language may seem nonsensical in another one, but it does not yet imply it is meaningless.

All the 2-bit probabilities we have considered in the paper, including the quantum ones,  have the same general form,
\be
p_l=g^l\left(\frac{\theta}{2\pi}\right)
=
g^l\left(\frac{g^{-l}\big(g^l(\theta)\big)}{g^{-l}\big(g^l(2\pi)\big)}\right)
=
g^l(\theta) \oslash^l g^l(2\pi)
=
\theta_l \oslash^l (2\pi)_l,
\label{141}
\ee
The case $l=1$ reconstructs quantum probabilities typical of the singlet state, provided we read the formulas by means of level-0 arithmetic,
\be
p_1=\frac{1}{2}\sin^2\frac{\theta}{2}.\label{142a}
\ee
If we read the same formulas by means of level-1 arithmetic, they appear completely classical,
\be
p_1=\theta_1 \oslash^1 (2\pi)_1,
\label{142}
\ee
expressing a ratio of two arc-lengths on a circle. We can say that (\ref{142a})--(\ref{142}) exemplify two different names for the same real number. The situation becomes even more clear if we choose, say, $\theta=\pi$. Then,
\be
p_1=\frac{1}{2}\sin^2\frac{\pi}{2}=\pi/(2\pi)=\pi_0 \oslash^0 (2\pi)_0=\pi_1 \oslash^1 (2\pi)_1
=\pi_2 \oslash^2 (2\pi)_2=\dots
\label{143}
\ee
Such a formula is as meaningful as,
\be
p_1=\frac{1}{2}\sin^2\frac{\pi}{2}=0.5=(0.5)_{10}=(0.1)_{2}=(0.8)_{16}=\dots
\label{144}
\ee
In principle, one can invent a situation where the inequality 
\be
0.1+0.1<1.0
\ee
is `violated' in experiment because one derives the inequality in the decimal system while the device that generates the data is based on binary coding (note that the numeral \citep{numeral} $1.0$ does represent the same number $1\in\mathbb{R}$ in both systems). 

The example illustrates the difference between the set $\mathbb{R}$ and decimal and binary arithmetics, 
$R_{10}=\{\mathbb{R},+_{10},-_{10},\cdot_{10},/_{10}\}$ and 
$R_{2}=\{\mathbb{R},+_{2},-_{2},\cdot_{2},/_{2}\}$.

Probabilities (\ref{141}) are `rotationally invariant' because the parameter $\theta$ is a difference of two parameters, $\alpha$ and $\beta$, associated with two experimental devices. Of course, the parameter does not have to correspond to a geometric angle of rotation. For example, it can represent a phase in an interferometer. 

The notion of difference is level-dependent. At the $l$th level one finds,
\be
\theta_l=g^l(\alpha-\beta)
=
g^l\Big(g^{-l}\big(g^l(\alpha)\big)-g^{-l}\big(g^l(\beta)\big)\Big)
=
g^l(\alpha)\ominus^l g^l(\beta)=
\alpha_l\ominus^l \beta_l.
\ee
The rotational invariance becomes explicit if one writes
\be
\alpha_l\ominus^l \beta_l
=
(\alpha_l\oplus^l\phi)\ominus^l (\beta_l\oplus^l\phi)
=
\alpha_l\oplus^l\phi\ominus^l\beta_l\ominus^l\phi
\ee
which is true for any $\phi\in R^l$. In general, if we denote $x_l=g^l(x)$, we observe that level-$l$ probabilities can be written as 
\be
p_l= |\alpha_l\ominus^l \beta_l|\oslash^l (2_l\odot^l \pi_l).
\ee
For any $l$ one can write a Clauser-Horne-type linear combination
\be
\theta_l\oslash^l (2_l\odot^l \pi_l)
\ominus^l
3_l\odot^l\theta_l\oslash^l (2_l\odot^l \pi_l)
\oplus^l
\theta_l\oslash^l (2_l\odot^l \pi_l)
\oplus^l
\theta_l\oslash^l (2_l\odot^l \pi_l)
=
0_l=0.\nonumber
\ee
In this sense, the {\it Clauser-Horne inequality is never violated\/}. What we know as the violation of the Clauser-Horne inequality by quantum probabilities is the following linear combination,
\be
\theta_1\oslash^1 (2_1\odot^1 \pi_1)
-
3_1\odot^1\theta_1\oslash^1 (2_1\odot^1 \pi_1)
+
\theta_1\oslash^1 (2_1\odot^1 \pi_1)
+
\theta_1\oslash^1 (2_1\odot^1 \pi_1)
,\nonumber
\ee
which equals $-0.20711$ if one takes $\theta=\pi/4$ (see the remark after Lemma~4).
Violation of the Clauser-Horne inequality reduces here to the trivial fact that
\be
3_1\odot^1\left(\theta_1\oslash^1 (2_1\odot^1 \pi_1)\right)
\neq
3 \cdot\left(\theta_1\oslash^1 (2_1\odot^1 \pi_1)\right).
\ee
It follows that violations of Clauser-Horne inequalities occur if one processes level-1 probabilities by the arithmetic rules of level 0. This is somewhat  like observing that $0.1+0.1=1.0$ in spite of the proof that $0.1+0.1<1.0$. 

The methodology of Bell's theorem is to prove that level-1 probabilities cannot result from a classical theory, where by the latter one means a theory based on level-0 calculus. However, exactly the same situation occurs if one takes any two neighboring levels of the hierarchy. So, the argument will apply to any $k$, even $k=0$ which we tend to  regard as the normal, standard, macroscopic,  human arithmetic. 

Level-($-1$) physicists can prove a Bell-type theorem stating that those who live at level 0 (that is, us)  do not exist as elements of reality. 
If level 1 is microscopic with respect to level 0, then level 0 is microscopic with respect to level $-1$.

It is conceptually refreshing to think of level $-1$ as a cosmic-scale arithmetic and calculus.  From such a perspective, we are small-scale microscopic objects whose existence is ruled out by logical and mathematical principles of galaxy-scale observers. Cosmic-scale arithmetic that occurs in the context of dark energy \citep{MCdark2} supports this interpretation. 

\section*{Acknowledgments}

Calculations were carried out at the Academic Computer Center in Gda{\'n}sk. 
The work was supported by the CI TASK grant `Non-Newtonian calculus with interdisciplinary applications'. 
KN has been supported by Ministry of Education and Science student fellowship `Szko{\l}a Or{\l}\'ow'.

\end{document}